\documentclass[12pt, draftclsnofoot, onecolumn]{IEEEtran}

\usepackage{amsfonts,amsmath,amssymb}
\usepackage{graphicx,color,epsfig,rotating}
\usepackage{latexsym}
\usepackage{subfigure}
\usepackage{cite}
\usepackage{epic,eepic}
\usepackage{algorithm}
\usepackage{algorithmic}
\usepackage{float}
\usepackage{multirow}
\usepackage{setspace}

\newtheorem{lemma}{Lemma}

\newfont{\bb}{msbm10 scaled 1100}

\newcommand{\RR}{\mbox{\bb R}}
\newcommand{\CC}{\mbox{\bb C}}

\newcommand{\Av}{{\bf{A}}}

\newcommand{\Jv}{{\bf{J}}}
\newcommand{\Iv}{{\bf{I}}}

\newcommand{\Sigmab}{{\boldsymbol{\Sigma}}}

\newcommand{\hv}{{\bf{h}}}
\newcommand{\ev}{{\bf{e}}}

\newcommand{\vv}{{\bf{v}}}

\newcommand{\Gc}{{\cal G}}

\newcommand{\Sc}{{\cal S}}
\newcommand{\Lc}{{\cal L}}

\newcommand{\herm}{{\sf H}}

\newcommand{\trace}{{\sf tr}}

\newcommand{\SINR}{{\sf sinr}}

\newcommand{\transp}{{\sf T}}

\newcommand{\CN}{\mathcal{CN}}

\newcommand{\be}{\begin{equation}}
\newcommand{\ee}{\end{equation}}
\newcommand{\bea}{\begin{eqnarray}}
\newcommand{\eea}{\end{eqnarray}}
\newcommand{\bitem}{\begin{itemize}}
\newcommand{\eitem}{\end{itemize}}
\newcommand{\benum}{\begin{enumerate}}
\newcommand{\eenum}{\end{enumerate}}

\begin{document}
\title{A New Energy Efficient Beamforming Strategy \\
for MISO Interfering Broadcast Channels\\
 based on  Large Systems Analysis}
\author{\IEEEauthorblockN{Sang-Rim Lee, Jaehoon Jung, Haewook Park, and Inkyu Lee, \textit{Senior Member, IEEE} \\}
\IEEEauthorblockA{School of Electrical Eng., Korea University, Seoul, Korea \\
              Email: \{sangrim78, jhnjung, jetaime01, inkyu\}@korea.ac.kr}
}\maketitle

\begin{abstract}
In this paper, we propose a new beamforming design to maximize energy efficiency (EE)
for multiple input single output interfering broadcast channels (IFBC).
Under this model,
the EE problem is non-convex in general due to the coupled interference and its fractional form, and thus it is difficult to solve the problem.
Conventional algorithms which address this problem have adopted an iterative method for each channel realization, which requires high computational complexity.
In order to reduce the computational complexity,
we parameterize the beamforming vector by scalar parameters related to beam direction and power.
Then, by employing asymptotic results of random matrix theory with this parametrization,
we identify the optimal parameters to maximize the EE
in the large system limit
assuming that the number of transmit antennas and users are large with a fixed ratio.
In the asymptotic regime,
our solutions depend only on the second order channel statistics,
which yields significantly reduced computational complexity and system overhead compared to the conventional approaches.
Hence, the beamforming vector to maximize the EE performance can be determined with local channel state information and the optimized parameters.
Based on the asymptotic results,
the proposed scheme can provide insights on the average EE performance, and
a simple yet efficient beamforming strategy is introduced for the finite system case.
Numerical results confirm that the proposed scheme shows a negligible performance loss compared to the best result achieved by the conventional approaches even with small system dimensions,
with much reduced system complexity.

\end{abstract}

\section{Introduction} \label{sec:intro}
%
A design of traditional wireless networks
focusing on high spectral efficiency
has caused rapidly increasing energy consumption and negative impact on the environment.
Therefore, pursuing high energy efficiency (EE)
becomes an important and urgent task for future wireless system designs \cite{Chen:11}.
In general, the EE is  defined as the ratio of the sum-rate  to the total power consumption measured in bit/Joule.
Meanwhile,
coordinated beamforming schemes,
which allow base stations (BSs) to jointly optimize their transmissions by sharing channel state information (CSI),
are considered as a key technology in cellular networks
due to its significant spectral efficiency improvement \cite{WBLee:13}.
When the EE is taken into account for a design of future wireless systems, cooperative transmission techniques need to be investigated with new perspectives.

From the EE point of view,
several papers have studied methods to maximize the performance various systems \cite{Isheden:12,HJKim:13,He:13,Kwan:12,Kwan:12_1,Bjornson:13,Ngo:13}.
The EE maximization problem in general belongs to a class of fractional programming due to its fractional form, and thus is nonlinear.
Nevertheless, for the special case of no interference among users,
the problem can be transformed into an equivalent convex problem
without loss of optimality by exploiting the pseudo concavity of the objective function \cite{Isheden:12}.
As a result, the global optimal solution can be found efficiently by convex optimization tools \cite{Boyd:04}.
Also, for the multi-user case,
the same framework can be adopted by employing zero-forcing beamforming \cite{Bjornson:15},
although the resultant performance is suboptimal in terms of EE.
However, in more general scenarios with inter-user interference,
the optimization problem for EE is non-convex, and thus
it is difficult and more challenging to optimize the EE in the presence of interference.

In this paper,
we focus on designing a new energy efficient scheme
for multi-cell multi-user downlink systems
where each BS equipped with multiple antennas communicates with its corresponding single-antenna users.
These systems can be mathematically modeled as multiple input single output (MISO) interfering broadcast channels (IFBC).
After transforming from fractional programming to linear programming in \cite{Isheden:12} and applying the weighted minimum mean square error (WMMSE) approach in \cite{Shi:11}, a local optimal solution was achieved in \cite{He:13}.
However, this method requires either centralized channel knowledge or the exchange of additional parameters.
In addition, the optimal beamforming vectors should be computed in an iterative manner for every channel realizations.
Moreover, it is difficult to get insights on average performance without resorting to Monte Carlo simulations.

To overcome these issues,
we propose a low complexity energy efficient scheme
with a negligible performance loss compared to the best results
achieved by the conventional approach.
To this end,
we first parameterize the beam vectors by the parameters associated with beam direction and power.
With this parameterization,
we then employ the asymptotic results of random matrix theory
\cite{Zakhour:13,Couillet:09,Huh:11,Hoydis:13}.
More specifically,
in the large system limit where the number of transmit antennas and users in each cell go to infinity with a fixed ratio,
we identify the parameters to optimize the EE.
Note that the beamforming vector which maximizes the EE performance is still constructed based on local instantaneous CSI in a finite dimension.
Meanwhile, the parameters can be optimized by adopting the large system analysis.
It is worth noting that
in the asymptotic regime,
the parameters
become deterministic and the randomness according to instantaneous channel realizations disappears.
Therefore,
only second order channel statistics is required for the large system approach.

In \cite{SRLee:13}, the sum rate (SR) maximization is performed for MIMO interference channel by utilizing the fact that a zero gradient value for sum rate maximization under fixed full power is efficiently obtained by utilizing a relationship between the SR and virtual signal-to-interference-plus-noise.
However, this method cannot be directly applied for the EE metric since beamforming direction and power allocation are jointly considered for the EE maximization.
Different from conventional EE algorithms which should be updated in each channel realization,
the proposed method does not recalculate the parameters as long as statistical channel information remains constant.
Thus, our long-term strategy significantly reduces complexity and the system overhead
compared to the conventional methods.
In addition, the dimensionality of the optimization problem
is greatly reduced by virtue of the asymptotic approach.
Moreover, an asymptotic expression of the achievable EE allows efficient evaluation of the system performance without the need of heavy Monte Carlo simulations.
The simulation results demonstrate that the
performance of the proposed scheme is almost identical to the
near-optimal EE  even for small system dimensions, with much reduced computational complexity.

The rest of this paper is organized as follows:
Section \ref{sec:sys_model} describes a system model and the problem formulation.
In Section \ref{sec:proposed_algo},
we present a low complexity beamforming design based on large system analysis,
and simulation results are presented in Section \ref{sec:simul}.
Finally, in Section \ref{sec:conclusions}, this paper is terminated with conclusions.

Throughout the paper,
we adopt uppercase boldface letters for matrices and lowercase boldface for vectors.
The superscripts $(\cdot)^{\transp}$ and $(\cdot)^{\herm}$ stand for transpose and  conjugate transpose, respectively.
In addition, $\|\cdot\|$, $\trace (\cdot)$, $[\cdot]_{k}$ and $[\cdot]_{ij}$ represent 2-norm, trace,
the $k$-th element of a vector and the \mbox{$(i,j)$-th} entry of a matrix, respectively.
Also, $\Iv_d$ denotes an identity matrix of size $d$ and $\bold{0}_d$ means a zero matrix of size $d \times d$.
A set of $N$ dimensional complex column vectors is defined by $\CC^N$
and $|\Sc|$ indicates the cardinality of the set $\Sc$.
\section{System Model} \label{sec:sys_model}

\begin{figure}
\centering
\includegraphics[width=4in]{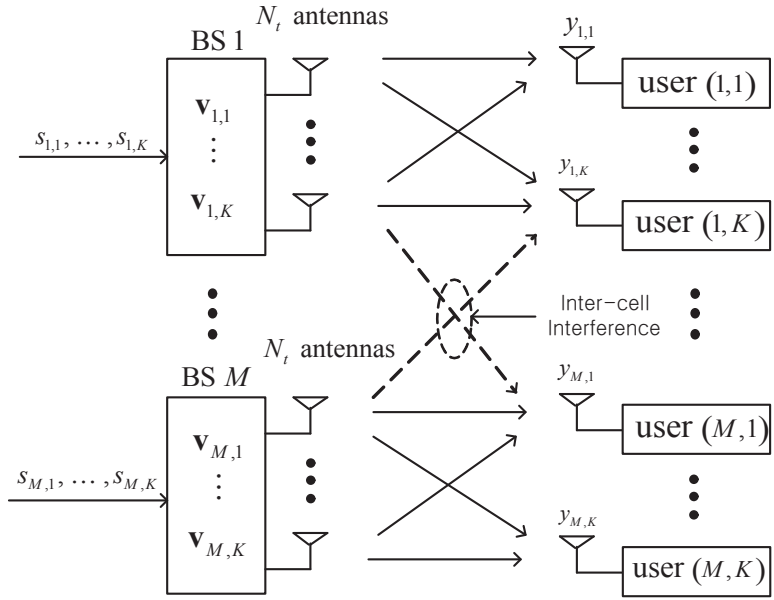}
\caption{The system model of $M$-cell MISO-IFBC}
\label{fig:sys_model}
\end{figure}

In this paper, we consider an $M$-cell MISO-IFBC with bandwidth $W_c$
where each BS equipped with $N_t$ transmit antennas serves $K$ users with a single antenna as shown in Fig. \ref{fig:sys_model}.
Here user $(j,k)$ indicates  the $k$-th user in \mbox{cell $j$}.
Denoting $\hv_{m,j,k} = \bold{R}_m^{\frac{1}{2}} \bold{z}_{m,j,k}$ where $\bold{R}_m\in\CC^{N_t \times N_t}$ is a deterministic transmit covariance matrix at the $m$-th BS and $\bold{z}_{m,j,k} \in \CC^{N_t \times 1}$ as the flat-fading channel vector from the $m$-th BS to the $k$-th user in cell $j$ with the coherence time $T_c$,
the received signal $y_{j,k}$ at user $(j,k)$ is expressed by
\bea
y_{j,k} = \hv_{j,j,k}^{\herm} \vv_{j,k} s_{j,k}
    + \sum_{(m,n) \neq (j,k)} \hv_{m,j,k}^{\herm} \vv_{m,n} s_{m,n} + n_{j,k} \nonumber
\eea
where  $\vv_{m,n} \in \CC^{N_t \times 1}$ equals the beamforming vector for user $(m,n)$,
$s_{m,n} \sim \CN(0,1)$  stands for the complex data symbol intended for user $(m,n)$,
and $n_{j,k} \sim \CN(0, \sigma^2)$ represents the additive white Gaussian noise at user $(j,k)$.
Throughout the paper, we assume that the entries of  $\hv_{m,j,k}$ are uncorrelated Rayleigh fading according to $\CN(0, \epsilon_{m,j,k})$
where $\epsilon_{m,j,k}$ indicates the pathloss from \mbox{BS $m$} to user $(j,k)$.
Note that our result can be easily extended to a more general channel model.
It is also assumed that $\sum_{k=1}^K \|\vv_{j,k}\|^2 \le P_j$ in order to satisfy  per-BS power constraint $P_j$.

We assume single user detection at the receiver so that each receiver treats  interference as the Gaussian noise.
Thus, the individual rate of user $(j,k)$ for given transmit beamforming vectors of all BSs $\{\vv_{m,n}\}$
is computed as
\bea
R_{j,k}(\{\vv_{m,n}\}) = \log_2 \left( 1 + \SINR_{j,k}\left(\{\vv_{m,n}\}\right)\right) \eea
where $\SINR_{j,k}(\{\vv_{m,n}\})$ represents the individual signal-to-interference-plus-noise ratio (SINR) for user $(j,k)$ expressed as
\bea
\SINR_{j,k}(\{\vv_{m,n}\}) = \frac{|\hv_{j,j,k}^{\herm} \vv_{j,k}|^2}{\sum_{(m,n) \neq (j,k)} |\hv_{m,j,k}^{\herm} \vv_{m,n}|^2 + \sigma^2}.
\eea
Then, the total amount of information transmitted during a time-frequency chunk $T_c W_c$ is given by
\bea
T_c W_c \sum_{j,k} \log_2(1+\SINR_{j,k}).~~~~[\text{bits}]\nonumber
\eea

Meanwhile, for designing an energy efficient transmission algorithm,
we consider the power consumption model for BSs in \cite{Kwan:12}.
Thus, the total energy consumption during the time-frequency block $T_c W_c$ is modeled as
\bea
P_T = T_c W_c\left( \zeta \sum_{j,k} \|\vv_{j,k}\|^2 + M N_t P_c + M P_0\right)~~~~[\text{Joule}]
\eea
where $\zeta \ge 1$ is a constant associated with the power amplifier inefficiency,
${P_c}$ is defined by ${P_c} = \frac{{P_c}^\prime}{W_c}$ with ${P_c}^\prime$ being the constant circuit power consumption
proportional to the number of radio frequency chains,
and ${P_0}$ denotes ${P_0} = \frac{{P_0}^\prime}{W_c}$ with ${P_0}^\prime$ accounting for the static power consumed at the BS which is independent of the number of transmit antennas.
For example, $P_c$ includes power
dissipation in the transmit filter, mixer, frequency synthesizer,
and digital-to-analog converter \cite{Kwan:12}.

Accordingly, the EE in bits/Joule
is defined as the weighted sum-rate (WSR) divided by the amount of the energy consumption as
\bea
\text{EE} = \frac{f_1(\{\vv_{j,k}\})}{f_2(\{\vv_{j,k}\})}
= \frac{\sum_{j,k} w_{j,k}R_{j,k}}{\zeta \sum_{j,k} \|\vv_{j,k}\|^2 + M N_t P_c + M P_0}. \eea
Here, it is assumed that the positive weight term $w_{j,k}$ is predetermined by a scheduler according to the priority of user $(j,k)$ \cite{He:13}.
Thus, the EE maximization problem can be formulated as
\bea \label{pro:orginal_ee}
& &\max_{\{\vv_{j,k}\}} \frac{f_1(\{\vv_{j,k}\})}{f_2(\{\vv_{j,k}\})}  \\
& &\mbox {s.t.} \sum_{k=1}^K \|\vv_{j,k}\|^2 \le P_j  \mbox{ for }  j = 1, \ldots, M. \nonumber
\eea

We notice that problem (\ref{pro:orginal_ee}) is in general non-convex due to coupled interference among users and its fractional form.
Therefore, identifying a solution of this problem is quite complicated.
As an alternative, by applying a transformation from the fractional programming into LP and the WMMSE approach sequentially,
a local optimal solution can be obtained as in \cite{He:13}.
%

\section{Conventional Approach for Energy Efficiency} \label{sec:proposed_algo}
In this section, we briefly review a conventional approach for the EE maximization in \cite{He:13}, which employs a two-layer optimization strategy.
First, in the outer layer, the fractional programming problem is transformed into a linear programming problem with a new parameter.
Then, for the given parameter, the inner problem is solved by using the WMMSE method developed in \cite{Shi:11}.
Eventually, a final solution is found by inner and outer loops iterations.

The optimization problem (\ref{pro:orginal_ee}) can be transformed into a linear programming problem
by introducing a new parameter $\eta$.
From the relationship between the fractional programming and the parametric programming \cite{Isheden:12},
the original problem (\ref{pro:orginal_ee}) can be recast as the following equivalent form
\bea \label{eq:frac_ee}
\max_{\{\vv_{j,k}\},~ \eta \in \RR^+}& & \eta \\
\mbox{s.t.}\quad& &{f_1(\{\vv_{j,k}\})} - \eta {f_2(\{\vv_{j,k}\})} \ge 0 \nonumber  \\
& &\sum_{k=1}^K \|\vv_{j,k}\|^2 \le P_j  \mbox{ for }  j = 1, \ldots, M. \nonumber
\eea

For a fixed value of $\eta$,
we have a feasibility problem in $\{\vv_{j,k}\}$
which checks if
$F(\eta) \ge 0$, where $F(\eta)$ indicates the optimal value of the following problem
\bea \label{pro:inner_pro}
\max_{\{\vv_{j,k}\}} & &
{f_1(\{\vv_{j,k}\})} - \eta {f_2(\{\vv_{j,k}\})} \\
\mbox{s.t.}& &\sum_{k=1}^K \|\vv_{j,k}\|^2 \le P_j  \mbox{ for }  j = 1, \ldots, M. \nonumber
\eea
From Theorem 1 in \cite{He:13},  $F(\eta)$ is shown to be a monotonically decreasing function with respect to $\eta$ and
the equation $F(\eta) = 0$ has a unique solution.
As a result, the optimal value of $\eta$ can be identified using one dimensional search algorithms such as a simple bisection method \cite{Boyd:04}.

Next, the optimal beamforming needs to be determined in the inner problem for a fixed $\eta$.
For the given $\eta$, the inner problem (\ref{pro:inner_pro}), excluding terms irrelevant to the optimization variables $\{\vv_{j,k}\}$, is rephrased as
\bea \label{pro:G}
\max_{\{\vv_{j,k}\}}& &
G(\{\vv_{j,k}\})  \\
\mbox{s.t.}& &  \sum_{k=1}^K \|\vv_{j,k}\|^2 \le P_j \quad \forall j \nonumber
\eea
where $G(\{\vv_{j,k}\}) = \sum_{j,k} (R_{j,k} - { \eta}\zeta \|\vv_{j,k}\|^2)$.

Note that this problem is quite similar to the SR maximization problem
except for the power term in the objective function.
Thus, using the relationship between the SR and the WMMSE,
a solution of the problem (\ref{pro:G}) can be computed from the following equivalent problem
\bea \label{pro:min_mse}
\min_{\{\vv_{j,k}\}, \{u_{j,k}\}, \{s_{j,k}\}}
& &\sum_{j,k} w_{j,k}(e_{j,k} s_{j,k} - \log_2 s_{j,k} - 1) + \eta \zeta \|\vv_{j,k}\|^2  \qquad \qquad  \\
\mbox{s.t.}\quad\quad & & \sum_{k=1}^K \|\vv_{j,k}\|^2 \le P_j \qquad \forall j \nonumber
\eea
where the mean square error $e_{j,k}$ is given by
\bea \label{eq:mse}
e_{j,k} = \left|u_{j,k} \hv_{j,j,k}^{\herm} \vv_{j,k} - 1\right|^2
+ \sum_{(m,n) \neq (j,k)} \left|u_{j,k} \hv_{m,j,k}^{\herm} \vv_{m,n}\right|^2
+ \left|u_{j,k}\right|^2 \sigma^2,
\eea
and $\{s_{j,k}\}$ and  $\{u_{j,k}\}$ are auxiliary variables.

The above problem is still non-convex in terms of $\{\vv_{j,k}\}, \{s_{j,k}\}, \{u_{j,k}\}$ jointly,
making the direct optimization of the problem difficult.
However, since the problem is convex with respect to each of the optimization variables $\{\vv_{j,k}\}, \{s_{j,k}\}$ and $\{u_{j,k}\}$,
we can solve the problem with one parameter by fixing the other two, i.e.,
the problem can be calculated by alternating the optimization method.
For {given $\{\vv_{j,k}\}$}, the optimal $u_{j,k}$ of the problem (\ref{pro:min_mse}) is obtained by
\bea \label{eq:opt_u}
u_{j,k}^{opt, \eta} = \frac{\hv_{j,j,k}^{\herm} \vv_{j,k}}{\sum_{m,n} |\hv_{m,j,k}^{\herm} \vv_{m,n}|^2 + \sigma^2}.
\eea
Furthermore, for fixed $\{u_{j,k}\}$ and $\{\vv_{j,k}\}$, the optimal $s_{j,k}$ is expressed by
\bea \label{eq:opt_s}
s_{j,k}^{opt, \eta} = \left(1 - u_{j,k} \vv_{j,j,k}^{\herm} \hv_{j,j,k}\right)^{-1}.
\eea

Then, once the values of $\{s_{j,k}\}$ and $\{u_{j,k}\}$ are given,
the optimization of $\{\vv_{j,k}\}$  is decoupled among the BSs by substituting $e_{j,k}$ in (\ref{eq:mse}) into the objective function of the problem (\ref{pro:min_mse}), and this leads to the following distributed optimization problems
for the $j$-th BS
\bea
\min_{\{\vv_{j,k}\}}& & \!\!\!\!\sum_{k=1}^K \left(w_{j,k}s_{j,k} |u_{j,k} \hv_{j,j,k}^{\herm} \vv_{j,k} - 1|^2 + \!\!\!\!\!\!\sum_{(m,n) \neq (j,k)}\!\!\!\!\!\!\!\! w_{m,n}s_{m,n} |u_{m,n}|^2 |\hv_{j,m,n}^{\herm} \vv_{j,k}|^2  + \eta \zeta \|\vv_{j,k}\|^2 \right) \\
\mbox{s.t.} & & \!\!\!\!\sum_{k=1}^K \|\vv_{j,k}\|^2 \le P_j. \nonumber
\eea

Denoting $\mu_j \ge 0$ as the Lagrange multiplier corresponding to the power constraint,  the first order optimality condition of the Lagrange function with respect to each $\vv_{j,k}$ yields
\bea  \label{eq:opt_v}
\vv_{j,k}^{opt, \eta} = w_{j,k} s_{j,k} u_{j,k} \left(  \sum_{(m,n)} w_{m,n}s_{m,n} |u_{m,n}|^2 \hv_{j,m,n} \hv_{j,m,n}^{\herm} + (\eta \zeta + \mu_j) \Iv \right)^{-1} \hv_{j,j,k}
\eea
where $\mu_j$ is  chosen such that the complementary slackness condition of power constraint is fulfilled. Let $\vv_{j,k} (\mu_j)$ be the right-hand side of (\ref{eq:opt_v}).
If $\sum_{k=1}^K  \|\vv_{j,k}(0)\|^2 \le P_j$, then $\mu_j^{opt, \eta} = 0$.
Otherwise, $\mu_j^{opt, \eta}$ can be found by using the bisection method
which satisfies $\sum_{k=1}^K  \|\vv_{j,k}(\mu_j)\|^2 = P_j$.
Therefore,
a solution of $\{\vv_{j,k}^{opt, \eta}\}$ for a given $\eta$
can be computed by updating $\{\vv_{j,k}\}, \{s_{j,k}\}$ and $\{u_{j,k}\}$ in an alternating fashion.

In summary,
a local optimal point of the EE can be determined by two-layer optimization.
However, the algorithm should be carried out in an iterative manner
per each instantaneous channel realization by sharing global channel information among BSs.
This leads to high computational complexity and signaling overhead.
In the following,
we will propose a new algorithm with low complexity and overhead
which is more desirable in practical systems.

\section{Proposed EE Scheme based on Large System Analysis} \label{sec:large_system}
In this section, we propose an energy efficient scheme with low complexity in a finite dimension.
After introducing conventional approaches, we describe the proposed method
based on the asymptotic results of random matrix theory.
Note that we consider the asymptotic regime where $N_t \rightarrow \infty$
with $\frac{K}{N_t}$ held at a fixed ratio to quantify beamforming parameters.
The key idea is to combine large system analysis techniques with the WMMSE approach.
Specifically,
by applying the equivalence property between SR and WMMSE,
the structure of the optimal beamforming vector is characterized
with parameters related to beam direction and power.
Then, by employing the asymptotic results of random matrix theory, the value of the parameters becomes deterministic which depends only on the second order channel statistics, and this leads to a significant reduction in the computational complexity compared to the system which utilizes instantaneous CSI.
As will be shown later,
the beamforming vector can be computed using only the optimized parameters and local CSI.

Before explaining the algorithm,
we provide useful results for solving the problem.
First, we identify the structure of the optimal beamforming based on $\vv_{j,k}^{opt, \eta}$ in (\ref{eq:opt_v}).
Note that from $\vv_{j,k}^{opt, \eta}$, the structure of beamforming can be paramterized with the power term $p_{j,k}$ and the parameters related to the beam direction $\beta_{j,k}$ and $\lambda_j$ as
\bea
\vv_{j,k}^{opt, \eta} &=& {\sqrt{p_{j,k}}} c_{j,k} {\left(\sum_{(m,n)} {\beta_{m,n}} \hv_{j,m,n} \hv_{j,m,n}^{\herm} + \lambda_j \Iv \right)^{-1}  \hv_{j,j,k}}
\eea
where $\beta_{j,k}$ and $\lambda_j$ are given as
\bea
\label{eq:beta_rel}
\beta_{m,n} &=& w_{m,n}s_{m,n} |u_{m,n}|^2, \\
\label{eq:lambda_rel}
\lambda_j &=& \eta \zeta + {\mu_j},
\eea
and  $p_{j,k}$ and $c_{j,k}$ are denoted as $p_{j,k} = \|\vv_{j,k}^{opt, \eta}\|^2$ and $c_{j,k} =  1/ \| (\sum_{(m,n)} {\beta_{m,n}} \hv_{j,m,n} \hv_{j,m,n}^{\herm} + \lambda_j \Iv )^{-1}  \hv_{j,j,k}\|$, respectively.
Based on this structure,
the normalized beam direction vectors  $\bar{\vv}_{j,k}$ is defined as
\bea \label{eq:norm_beam}
\bar{\vv}_{j,k}
= c_{j,k} {\left( \sum_{(m,n)} \beta_{m,n} \hv_{j,m,n} \hv_{j,m,n}^{\herm} + \lambda_j \Iv \right)^{-1} \hv_{j,j,k}}
\eea
where $\beta_{m,n}$ and $\lambda_j$ represent
the parameters which control the leakage interference power level to other users adaptively.

In order to quantify the component-wise impact on the performance for the given beamforming vectors $\{\bar{\vv}_{j,k}\}$,
we introduce the normalized channel gain matrix as
\bea
\Gc =
\begin{bmatrix}
        |\hv_{1,1,1}^{\herm} \bar{\vv}_{1,1}|^2 & |\hv_{1,1,2}^{\herm} \bar{\vv}_{1,1}|^2 & \cdots & |\hv_{1,M,K}^{\herm} \bar{\vv}_{1,1}|^2 \\
        |\hv_{1,1,1}^{\herm} \bar{\vv}_{1,2}|^2 & |\hv_{1,1,2}^{\herm} \bar{\vv}_{1,2}|^2 & \cdots & |\hv_{1,M,K}^{\herm} \bar{\vv}_{1,2}|^2 \\
        \vdots & \vdots & \ddots & \vdots \\
        |\hv_{M,1,1}^{\herm} \bar{\vv}_{M,K}|^2 & |\hv_{M,1,2}^{\herm} \bar{\vv}_{M,K}|^2 & \cdots & |\hv_{M,M,K}^{\herm} \bar{\vv}_{M,K}|^2 \\
      \end{bmatrix} \in \RR^{M K \times M K}.
\eea

As mentioned before, in order to compute these instantaneous channel gains in a finite dimension, we exploit the results of the RMT for an asymptotic region.
The $((j-1) \times K + k, (m-1) \times K + n)$-th off-diagonal element of $\Gc$ accounts for interference power at user $(m,n)$
generated by the $j$-th BS for serving its supporting user $(j,k)$,
and the diagonal elements stand for the desired signal power.
Employing the asymptotic results of random matrix theory,
we arrive at the following lemma.
\begin{lemma} \label{lem:G}
For fixed ratio $\frac{K}{N_t}$ with $N_t \rightarrow \infty$, the deterministic equivalent of $\Gc$ is obtained as
\bea
\Gc - \Gc^{\circ} \xrightarrow{ a.s.} \bold{0}_{MK},
\eea
where
\bea
\Gc^{\circ} =
\begin{bmatrix}
          D_{1,1}^{\circ} & I_{1,1,1,2}^{\circ} & \cdots & I_{1,1,M,K}^{\circ} \\
          I_{1,2,1,1}^{\circ} & D_{1,2}^{\circ} & \cdots & I_{1,2,M,K}^{\circ} \\
          \vdots & \vdots & \ddots & \vdots \\
          I_{M,K,1,1}^{\circ} & I_{M,K,1,2}^{\circ} & \cdots & D_{M,K}^{\circ} \\
\end{bmatrix} \in \RR^{M K \times M K }.
\eea
Here, $D_{j,k}^{\circ}$ and $I_{j,k,m,n}^{\circ}$ are given by
\bea
D_{j,k}^{\circ} &=& \frac{(m_{j,k}^{\circ})^2}{N_t \Psi_{j,k}^{\circ}}, \\
I_{j,k,m,n}^{\circ} &=& \frac{\Psi_{j,k,m,n}^{\circ}}{\left(1 + \beta_{m,n} m_{j,k,m,n}^{\circ}\right) \Psi_{j,k}^{\circ}},
\eea
where $m_{j,k}^{\circ}, m_{j,k,m,n}^{\circ}, \Psi_{j,k}^{\circ}$ and $ \Psi_{j,k,m,n}^{\circ}$ are provided in the proof.
\end{lemma}
\begin{IEEEproof}
See Appendix \ref{app:B}.
\end{IEEEproof}

With $\Gc^{\circ}$,
the deterministic equivalent of $\SINR_{j,k}$ is derived as
\bea
\SINR_{j,k} - \SINR_{j,k}^{\circ} \xrightarrow{N_t \rightarrow \infty} 0,
\eea
where
\bea
\SINR_{j,k}^{\circ} = \frac{g_{j,k,j,k}^{\circ} p_{j,k}}{\sum_{(m,n) \neq (j,k)} g_{m,n,j,k}^{\circ} p_{m,n} + \sigma^2}.
\eea
Here, $g_{m,n,j,k}^{\circ}$ represents the $(m \times n, j \times k)$-th element of $\Gc^{\circ}$.
By the continuous mapping theorem \cite{Billingsley:95}, one can show that
the deterministic equivalents of the SR and the EE are expressed as
\bea
R_{\Sigma}^{\circ} &=& \sum_{j,k} w_{j,k}\log_2\left(1 + \frac{g_{j,k,j,k}^{\circ} p_{j,k}}{\sum_{(m,n) \neq (j,k)} g_{m,n,j,k}^{\circ} p_{m,n} + \sigma^2} \right) \\
\eta^{\circ} &=& \frac{R_{\Sigma}^{\circ}}{\zeta \sum_{j,k} p_{j,k} + M N_t P_c + M P_0}.
\eea

In what follows,
based on the above asymptotic results,
we optimize the EE performance in the large system limit instead of the original problem (\ref{pro:orginal_ee}).
The EE problem in the asymptotic regime
is formulated as
\bea
\max_{\{p_{j,k}\} ,\{\beta_{j,k}\}, \{\lambda_{j}\}  } & & \eta^{\circ}  \\
\mbox{s.t.\qquad} & & \sum_{k=1}^K p_{j,k} \le P_j \qquad \forall j. \nonumber
\eea

In the outer layer optimization,
our algorithm is analogous to the conventional approach shown in \mbox{Section III}.
However, unlike the conventional approach adopting a short-term strategy,
we consider a long-term strategy in order to achieve low complexity.
Notice that for the finite dimensional case,
the outer layer algorithm is required only to generate $\eta$ and pass to the inner problem.
Then, $\eta$ is updated according to the feasibility, i.e., $F(\eta) \ge 0$ or not, based on a solution of the inner problem.
Similar to the finite case, the feasibility in the asymptotic regime can be determined whether $F^{\circ}(\eta) \ge 0$ or not, where $F^{\circ}(\eta)$ is the optimal value of the following problem
\bea \label{pro:F_det}
\max_{\{p_{j,k}\} ,\{\beta_{j,k}\}, \{\lambda_{j}\} }& &  R_{\Sigma}^{\circ} - \eta \left( \zeta \sum_{j,k} p_{j,k} + \sum_{j}(N_t P_c + P_0)\right)\\
\mbox{s.t.\qquad} & & \sum_{k=1}^K p_{j,k} \le P_j \qquad \forall j. \nonumber
\eea

Now, the only remaining work in the outer layer optimization is to identify the maximum value of $\eta$ for a bisection method.
It is clear that the maximum value can be obtained as
\bea
\eta_{max} = \frac{\sum_{j,k} w_{j,k}\log_2(1 + \frac{P_j}{\sigma^2}\| \hv_{j,j,k}\|^2)}{\sum_j (N_t P_c + P_0)}.
\eea
Thus, applying Theorem 3.4 in \cite{Debbah:11},
it follows
\bea \label{eq:eta_max}
\eta_{max}^{\circ} = \frac{\sum_{j,k} w_{j,k}\log_2(1+ \frac{P_j N_t }{\sigma^2})}{\sum_j (N_t P_c + P_0)}
\eea
where we have used $\|\hv_{j,j,k}\|^2 - N_t \xrightarrow{a.s.} 0$.

Next, we derive the optimal beamforming in the inner problem for a fixed $\eta$.
Similar to (\ref{pro:G}),
the objective function in the problem (\ref{pro:F_det}) can be reduced to
\bea \label{eq:G_asym}
G^{\circ}(\{\beta_{j,k}\},\{\lambda_j\}, \{p_{j,k}\}) = R_{\Sigma}^{\circ} - { \eta}\zeta\sum_{j,k} p_{j,k}.
\eea
Then, by utilizing the WMMSE approach,
the problem (\ref{pro:F_det}) can be recast as
\bea \label{eq:mse_min_det}
\max_{\{p_{j,k}\} ,\{\beta_{j,k}\}, \{\lambda_{j}\}, \{u_{j,k}\}, \{s_{j,k}\}}
 & & \sum_{j,k} (w_{j,k}(\tilde{e}_{j,k} - \log_2 s_{j,k} -1) + \eta\, \zeta\, p_{j,k})  \qquad \qquad \\
\mbox{s.t.\qquad\qquad} & & \sum_{k=1}^K p_{j,k} \le P_j \qquad \forall j \nonumber
\eea
where the mean square error $\tilde{e}_{j,k}$ is given by
\bea
\tilde{e}_{j,k} = \left(u_{j,k} \sqrt{g_{j,k,j,k}^{\circ} p_{j,k}} - 1\right)^2
+ \sum_{(m,n) \neq (j,k)} u_{j,k}^2 g_{m,n,j,k}^{\circ} p_{m,n}
+ u_{j,k}^2 \sigma^2. \nonumber
\eea

Thus, for any given $\{p_{j,k}\} ,\{\beta_{j,k}\}, \{\lambda_{j}\}$, the optimal receiver filters of the problem (\ref{eq:mse_min_det}) are obtained by
\bea \label{eq:opt_u_det}
u_{j,k}^{opt} = \frac{\sqrt{g_{j,k,j,k}^{\circ} p_{m,k}}}{\sum_{m,n} g_{m,n,j,k}^{\circ} p_{m,n} + \sigma^2}.
\eea
Furthermore, the optimal $s_{j,k}$ is expressed by
\bea \label{eq:opt_s_det}
s_{j,k}^{opt} = \frac{1}{ 1- u_{j,k}\sqrt{g_{j,k,j,k}^{\circ} p_{j,k}}}.
\eea
It is obvious  that
the optimal $\beta_{j,k}^{opt}$ in (\ref{eq:mse_min_det}) is equal to $s_{j,k}^{opt} (u_{j,k}^{opt})^2$ from (\ref{eq:beta_rel}) for the finite case.
Thus, we can calculate new $\Gc^{\circ}$ based on the updated $\{\beta_{j,k}^{opt}\}$.

Next, for given $\{s_{j,k}\}$ and $\{u_{j,k}\}$, the distributed problem for the $j$-th BS in the large system regime becomes
\bea
 \min_{\{p_{j,k}\}, \{\lambda_j\}}& &  \!\!\!\!\sum_{k=1}^K \left(w_{j,k}s_{j,k} |u_{j,k} \sqrt{ p_{j,k}  g_{j,k,j,k}^{\circ}} - 1|^2 + \!\!\!\!\sum_{(m,n) \neq (j,k)} \!\!\!\!\!\!w_{m,n}s_{m,n} |u_{m,n}|^2 g_{m,n,j,k}^{\circ}  + \eta \zeta p_{j,k} \right)\\
 \mbox{s.t.}\quad & & \!\!\!\!\sum_{k=1}^K p_{j,k} \le P_j \nonumber.
\eea
Then, the optimal transmit power $\{p_{j,k}\}$ is written by
\bea  \label{eq:opt_p_det}
p_{j,k}^{opt} = \frac{w_{j,k} s_{j,k} u_{j,k} \sqrt{g_{j,k,j,k}^{\circ}}}{\sum_{m,n} w_{m,n} s_{m,n} u_{m,n}^2 g_{j,k,m,n}^{\circ} + \lambda_j^{opt}}.
\eea
Here, from (\ref{eq:lambda_rel}),
$\lambda_j^{opt}$ is equal to $\lambda_j^{opt} = \eta \zeta + \mu_j^{opt}$
where $\mu_j^{opt}$ is determined by the complementary slackness of power constraint.

Let us denote $p_{j,k}(\mu_j)$ as the right-hand side of (\ref{eq:opt_p_det}).
If $\sum_{k=1}^K p_{j,k}(0) \le P_j$, then $p_{j,k}^{opt} =  p_{j,k}(0)$.
Otherwise, we must have
\bea \label{eq:pw_const}
\varphi(\mu_j) = \sum_{k=1}^K p_{j,k}(\mu_j) = P_j.
\eea
According to the monotonic property of the function $\varphi(\mu_j)$ with respect to $\mu_j$, equation (\ref{eq:pw_const}) can be efficiently solved by a bisection method.

\subsection{Overall Algorithm and Complexity Analysis}

The Algorithm \ref{alg:out} and \ref{alg:in} describe
the overall procedure of the proposed scheme.
Here, $\delta$ indicates a predefined threshold.
It is worth noting that our proposed algorithm
depends only on the second channel statistics,
and not on instantaneous channel realizations.
More specifically,
the conventional method determines $\{\beta_{j,k}\},\{\lambda_j\}$
and $\{p_{j,k}\}$ per each channel realization,
while the proposed algorithm does so only when the second order statistics changes, i.e., signal-to-noise ratio (SNR) changes.
Once the optimal $\{\beta_{j,k}\},\{\lambda_j\}$
and $\{p_{j,k}\}$ are determined,
we can construct the beamforming vectors based only on local CSI without additional complexity.
Thus, the proposed algorithm dramatically reduces  the computational complexity compared to the conventional scheme.

\begin{algorithm}[t]
\caption{Outer Layer} \label{alg:out}
\begin{algorithmic}[1]
\vspace{1mm}
\STATE Initialize $\eta_{min} = 0$ and $\eta_{max} = \eta_{max}^{\circ}$ in (\ref{eq:eta_max})\vspace{1mm}

\WHILE{$|\eta_{max} - \eta_{min}| > \delta$}
\vspace{1mm}

\STATE{ $\eta = \frac{\eta_{min} + \eta_{max}}{2}$}
\vspace{1mm}

\STATE{ Obtain the optimal solutions $\{\beta_{j,k}\}, \{\lambda_j\}, \{p_{j,k}\}$ by Algorithm \ref{alg:in}}
\vspace{1mm}

\STATE{ Compute $F^{\circ}(\eta)$}
\vspace{1mm}

\IF{$F^{\circ}(\eta) \le 0$}
\STATE{ $\eta_{max} = \eta$}
\vspace{1mm}
\ELSE
\STATE {$\eta_{min} = \eta$}
\vspace{1mm}
\ENDIF

\ENDWHILE
\end{algorithmic}
\end{algorithm}

\begin{algorithm}[t]
\caption{Inner Layer}
\label{alg:in}
\begin{algorithmic}[1]
\vspace{1mm}
\STATE Initialize $n=1$, $\beta_{j,k}^{(n)} = w_{j,k}$,  $\lambda_j^{(n)} = \eta \zeta, p_{j,k}^{(n)} = P_j/K$  for all   $j,k$
\vspace{1.5mm}

\STATE Set $G^{\circ}(\{\beta_{j,k}^{(0)}\},\{\lambda_j^{(0)}\},\{p_{j,k}^{(0)}\}) = 0$
\vspace{1.5mm}


\STATE Compute $\Gc^{\circ}$
\vspace{1.5mm}

\WHILE{$|G^{\circ}(\{\beta_{j,k}^{(n)}\},\{\lambda_j^{(n)}\},\{p_{j,k}^{(n)}\}) - G^{\circ}(\{\beta_{j,k}^{(n-1)}\},\{\lambda_j^{(n-1)}\},\{p_{j,k}^{(n-1)}\})| > \delta $}
\vspace{1.5mm}
\STATE $n \leftarrow n +1$
\vspace{1.5mm}

\STATE Update $u_{j,k}^{(n)}, s_{j,k}^{(n)}$ with (\ref{eq:opt_u_det}) and (\ref{eq:opt_s_det})  for all $j,k$
\vspace{1.5mm}

\STATE $\beta_{j,k}^{(n)} = w_{j,k}(u_{j,k}^{(n)})^2 s_{j,k}^{(n)}$
\vspace{1.5mm}

\STATE Calculate $\Gc^{\circ}$
\vspace{1.5mm}

\STATE Compute $p_{j,k}^{(n)}, \lambda_j^{(n)}$ with (\ref{eq:opt_p_det}) for all $j,k$
\vspace{1.5mm}

\ENDWHILE


\end{algorithmic}
\end{algorithm}

In what follows,
we compare the complexity of the conventional scheme with that of the proposed method.
For comparison,
the overall computational complexity
can be characterized by
the multiplication of the following terms:
the execution rate of the algorithm,
the iteration number of the outer layer,
the iteration number of the inner layer,
and the complexity of inner layer optimization per each iteration.
Here, the per-iteration complexity of the outer layer optimization is ignored
since the calculation using the bisection method
is relatively simple.

First, the per-iteration computational complexities
of the conventional scheme and the proposed method
are $\mathcal{O}(M^2 K^2 N_t^3)$ and $\mathcal{O}(M^2 K^2)$, respectively.
The difference between the two schemes
comes from an inverse operation of an $N_t \times N_t$ matrix.
The conventional algorithm requires the inverse operation to generate the beamforming vectors $\{\vv_{j,k}\}$ in (\ref{eq:opt_v}),
while the proposed  scheme does not need as can be seen in Algorithm \ref{alg:in}.
Thus, a complexity gain becomes larger as $N_t$ increases.
Next, for the inner and outer layer algorithm,
the required number of iterations of both schemes are quite similar in average sense,
since both of them are based on the bisection method and the WMMSE algorithm.

Compared to the conventional scheme,
the factor which reduces the complexity the most in the proposed algorithm is the execution rate of the overall algorithm.
The update rate of the proposed algorithm depends on how often the second order channel statistics changes, and thus is much slower than that of the conventional method
which needs to update at each realization.
This is because
large scale fading varies with tens of seconds, while small scale fading changes with few milliseconds in general wireless environments \cite{Tse:05}.
Thus, the coherence time of small scale fading is typically 1000 times smaller than that of large scale fading.
As a result, for example, with $M = 3$, $N_t = 4$, and $K = 3$, the CPU running time of the conventional EE algorithm is about 300 times more than that of the proposed scheme.
Therefore, we can verify that the proposed algorithm greatly reduces the computational complexity compared to the conventional scheme.
We shall show in the simulation section that our proposed algorithm exhibits the performance almost identical to that of the conventional algorithm while requiring significantly reduced complexity.


%
%
%
\section{Numerical results} \label{sec:simul}
\begin{center}
\begin{table}[t]
\centering
\caption{System setup}\label{tab:sys_config}
\begin{tabular}{|l|c|}
\hline
    System bandwidth ($\mathcal{W}$) & 20 MHz \\ \hline
	The number of user drops  & 10 \\ \hline
	The number of channel realizations per user drop & 100 \\ \hline
  The number of Tx antennas for each BS $N_t$ & 4 \\ \hline
  Cell radius $R$ & 500 m \\ \hline
  Minimum distance from BS to each user $R_{min}$ & 35 m \\  \hline
	Pass loss exponent $\alpha$ & 3.8 \\ \hline
	Transmit power constraint per BS $P_j$ & 26 $\sim$ 46 dBm \\ \hline
  Circuit power per antenna  $P_c$ & 30 dBm \\  \hline
  Basic power consumed at BS  $P_0$ & 40 dBm \\  \hline
  Noise figure ($N_F$) & 7 dB\\ \hline
  Noise power $\sigma^2$   &  -94 dBm  \\  \hline
  Inefficiency of the power amplifier $\zeta$ & 2 \\ \hline
\end{tabular}
\end{table}
\end{center}

In this section,
we evaluate the EE performance of the proposed beamforming scheme.
We consider a cooperative cluster of $M = 3$ hexagonal cells
for Monte Carlo simulations.
These simulations are carried out with the parameters listed in Table \ref{tab:sys_config}, unless specified otherwise.
The pathloss from BS $m$ to user $(j,k)$ $\epsilon_{m,j,k}$
is given as $10 \log_{10}\epsilon_{m,j,k} = -38 \log_{10} d_{m,j,k}- 34.5 $ in decibels,
where $d_{m,j,k}$ in meter indicates the distance from BS $m$ to user $(j,k)$.
Also, the noise power can be calculated as $\sigma^2 = -174 + 10\log(\mathcal{W}) + N_F$ in dBm where $\mathcal{W}$ means the system bandwidth and $N_F$ denotes the noise figure.

\begin{figure}\centering
\includegraphics[width=5.5in]{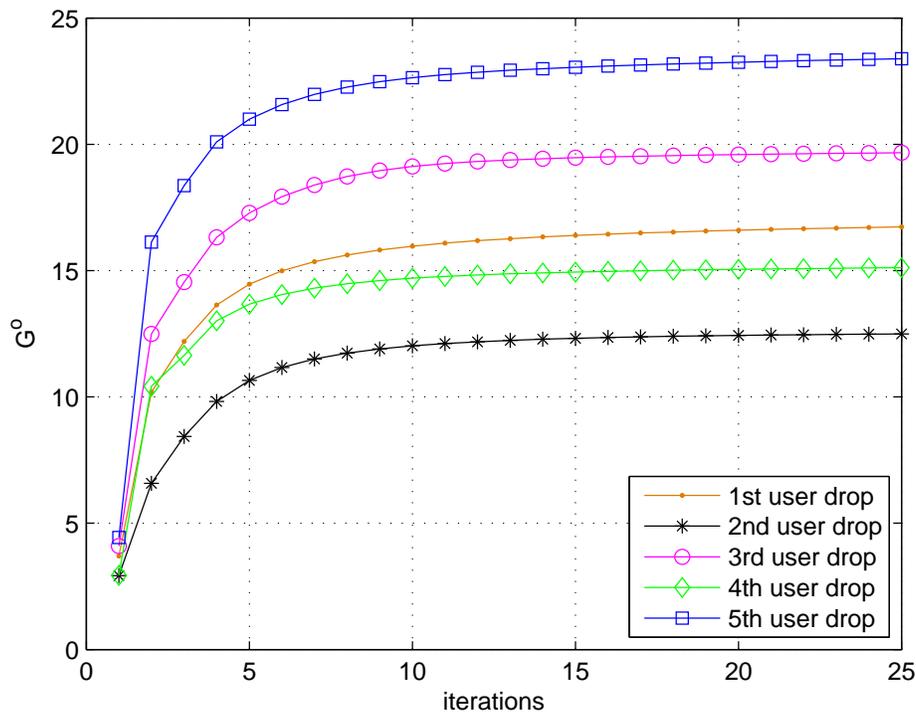}
\caption{Convergence examples of the inner layer algorithm in the proposed scheme}
\label{fig:conv_G}
\end{figure}

First, we illustrate the convergence of the proposed algorithm.
For the case of the outer layer optimization,
the optimal $\eta$ can be found based on one dimensional line search without loss of optimality, and thus the convergence is guaranteed.
On the other hand,
the inner layer algorithm cannot achieve the global optimal value due to the non-convexity of the problem.
However, the convergence to a local optimal point is guaranteed
by virtue of the WMMSE approach \cite{Solodov:98}.
Fig. \ref{fig:conv_G} plots the objective function $G^{\circ}$ in (\ref{eq:G_asym}) with respect to the number of iterations for $K = 4$.
The convergence trend varies with parameters such as power, user position and $\eta$.
In this figure, the curves corresponding to 5 different user drop events are plotted
by fixing certain $\eta$ and $P_j = 46$ dBm.
As shown in this plot,
the inner layer algorithm converges to a stable point with about 10 iterations.

\begin{figure}[t]\centering
\includegraphics[width=5.5in]{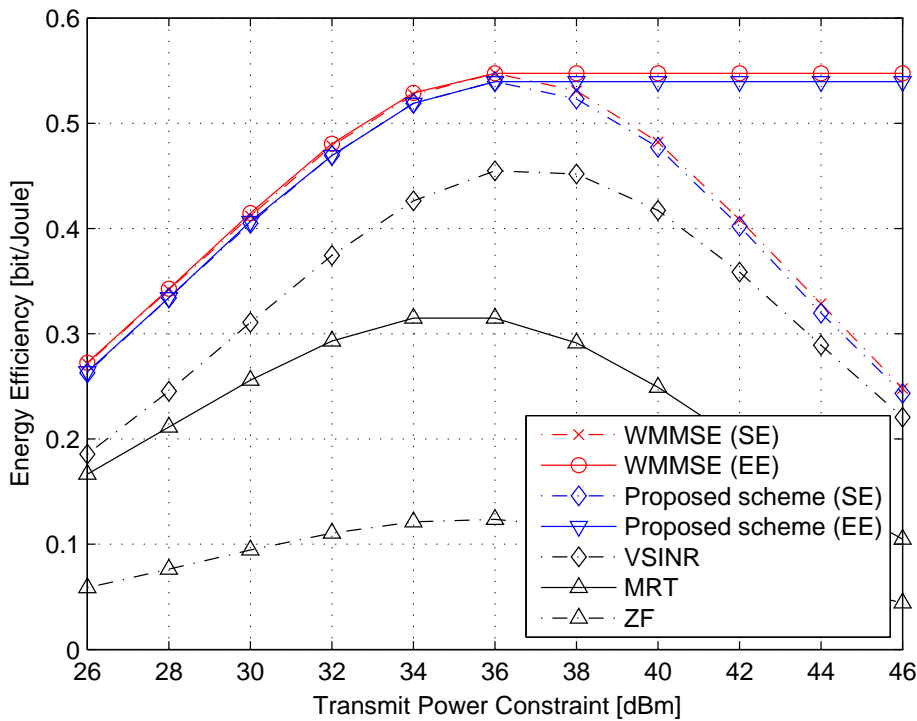}
\caption{Average EE performance of various beamforming strategies with $K = 3$}
\label{fig:EE_ref}
\end{figure}

Fig. \ref{fig:EE_ref} exhibits the average EE performance of various beamforming schemes as a function of $P_j$ for $K = 3$ with $w = [1~2~3]$.
For comparison,
we first present the EE performance of the following beamforming schemes.
\bitem
\item Maximal ratio transmission (MRT): the beamformers are aligned with the corresponding channels.
\item Zero-forcing beamforming (ZFBF): the signal to unintended users is nullified.
\item Conventional virtual SINR (VSINR): the VSINR is maximized with non-weighted coefficients,
i.e., $\beta_{j,k} = 1$ and $\lambda_j = \frac{\sigma^2}{P_j}$ for all $i,j$ \cite{Zakhour:10, SHPark:12}.
\item WMMSE algorithm: beamformers are designed to maximize the SR by using the WMMSE approach \cite{Shi:11}.
\item Conventional EE algorithm: the algorithm based on the WMMSE approach is adopted to maximize EE as described in Section III.
\item Proposed EE algorithm: the proposed algorithm performs with adaptive control of $\{\beta_{j,k}\}, \{\lambda_j\}$ and $\{p_{j,k}\}$ based on second order channel statistics for the EE maximization.
\eitem
For both the WMMSE and the conventional EE schemes which achieve the local optimal solution,
a solution of the VSINR scheme is adopted as an initial point.
Surprisingly,
we can see that the proposed scheme achieves near-optimal performance with much reduced complexity for all simulated transmit power constraint ranges.
It is emphasized again that
our proposed algorithm is performed only when second order channel statistics changes
and the constant values of $\{\beta_{j,k}\}, \{\lambda_{j}\}$ and $\{p_{j,k}\}$ are employed for generating beamforming vectors as long as the statistics remains unchanged.
This results in a significant computational complexity reduction compared to the conventional EE scheme which should be carried out in every channel realizations.
Moreover, we can observe the trade-off relationship between the performance and the complexity for various beamforming strategies.
Simple beamforming schemes such as MRT, ZFBF, and VSINR require lower computational complexity to comprise the beamforming structure than the proposed scheme but with poor performance.
In the ZFBF case, the EE performance is mainly degraded by the deficiency of dimension for nullifying the unintended user signals.
We also observe that
the WMMSE schemes designed for spectral efficiency maximization
produce much worse EE performance compared to the proposed scheme.
Especially,
a performance gain of the proposed EE scheme is about $209\%$ at $P_j = 46$ dBm $\forall j$.
\begin{figure}\centering
\includegraphics[width=5.5in]{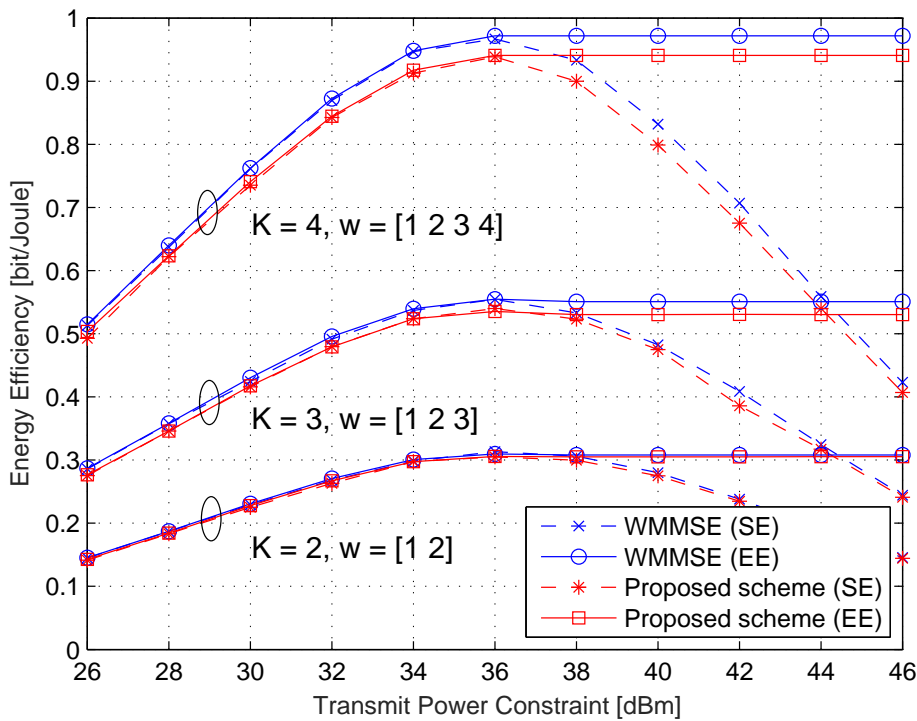}
\caption{The EE performance for different number of users}
\label{fig:EE_234}
\end{figure}

Also, the average EE performance for the conventional and proposed EE algorithms is illustrated for various number of users $K$ in Fig. \ref{fig:EE_234}.
We can see that the EE performance of these two algorithms increases with the number of users.
Moreover, the performance of the proposed EE algorithm has a small gap compared to the conventional algorithm, which is less than $4\%$ for all cases.
%

\begin{figure}\centering
\includegraphics[width=5.5in]{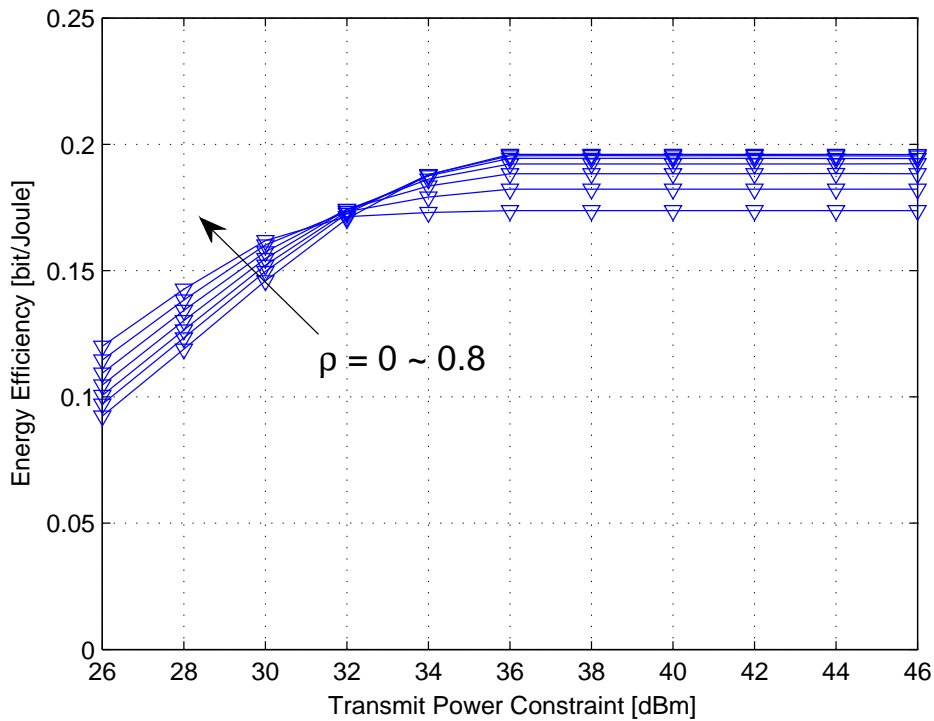}
\caption{The EE performance with transmit antenna correlation (K = 2)}
\label{fig:EE_corr}
\end{figure}
In Fig. \ref{fig:EE_corr}, we demonstrate the EE performance of the proposed scheme for the correlated transmit antenna case.
The transmit covariance matrix is set as the exponential correlation model which is given by $[\bold{R}]_{ij} = \rho^{|i-j|}$ with $i,j = 1, \cdots, N_t$ and $\rho \in [0,1)$.
In this correlation model, the average EE performance is enhanced when the correlation coefficient $\rho$ grows at a high transmit power region.
On the contrary, the opposite trend is observed at a low transmit power region.
This is due to the fact that the EE performance can be mainly affected by an array gain at the low power region.
For the high BS power region, the effect of spatial multiplexing takes a key role of the EE performance.

\begin{figure}\centering
\includegraphics[width=5.5in]{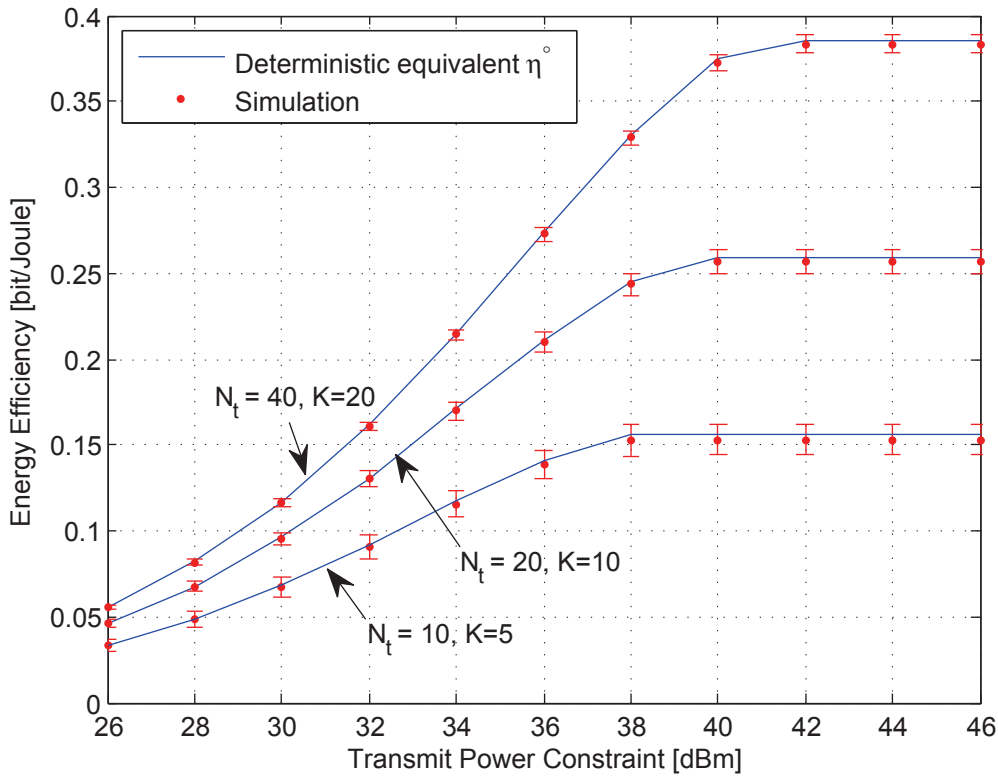}
\caption{Comparison between the average EE and the deterministic equivalent for $\frac{N_t}{K} = 2$}
\label{fig:EE_asym}
\end{figure}

In what follows,
we validate the accuracy of the deterministic equivalent of EE compared to true EE.
Fig. \ref{fig:EE_asym} compares the average EE with the deterministic approximation $\eta^{\circ}$ for $N_t = 10, 20$ and $40$ with the fixed ratio $\frac{N_t}{K} = 2$.
In this plot,
each curve corresponds to a particular drop of users for each $N_t$.
The error bars indicate the standard deviation of the simulation results.
As shown in this plot,
the deterministic equivalent of EE
provides a very accurate approximation.
It can be seen that the approximation lies within one standard deviation of
the Monte Carlo simulations and the standard deviation becomes smaller as $N_t$ increases.
Also,
we can check
that the maximum value of EE gets larger as $N_t$ grows.
This comes from an increased multiplexing gain, (i.e. pre-log term) as $N_t$ and $K$ grow larger.
From the plots,
it is observed that the EE performance curve increases up to a certain point, and after that it is saturated.
Investigation of the saturation point will be an interesting future work.


%
%
%
%
%
%
\section{Conclusions} \label{sec:conclusions}
In this paper,
we have proposed a low complexity beamforming scheme for MISO-IFBC.
With the parametrization of the beamforming vectors by the scalar values,
we have found the optimal parameters to maximize the EE in the asymptotic regime.
Our solutions
depend only on second order channel statistics, not on instantaneous CSI,
and thus
the parameters are computed only when channel
statistics changes. As a result, the computational complexity
is significantly reduced compared to the conventional method.
Through simulations, we have confirmed that the proposed
schemes with the asymptotic results provide  the near-optimal EE performance even for the finite system case.
Additionally,
the proposed scheme allows efficient calculation of the system performance without resorting to heavy Monte Carlo simulations.

\appendices

\section{Proof of Lemma \ref{lem:G} } \label{app:B}
We will derive the deterministic equivalents
of the desired signal power $|\hv_{j,j,k}^{\herm} \bar{\vv}_{j,k}|^2$
and the interference power $|\hv_{j,m,n}^{\herm} \bar{\vv}_{j,k}|^2$ subsequently as in \cite{SRLee:13}.
For simplicity, we assume $\bold{R}_i = \bold{R}$.

\subsubsection{Deterministic equivalent for $|\hv_{j,j,k}^{\herm} \bar{\vv}_{j,k}|^2$}
For given $\{\beta_{j,k}\}$ and $\{\lambda_j\}$,
$|\hv_{j,j,k}^{\herm} \bar{\vv}_{j,k}|^2$ is written by
\bea
|\hv_{j,j,k}^{\herm} \bar{\vv}_{j,k}|^2
&=& \frac{|\hv_{j,j,k}^{\herm} \left(\Av_j + \lambda_j \Iv\right)^{-1} \hv_{j,j,k}|^2}{\|\left(\Av_j + \lambda_j \Iv\right)^{-1} \hv_{j,j,k}\|^2}  \nonumber \\
&=& \frac{|\hv_{j,j,k}^{\herm} \left(\Av_j + \lambda_j \Iv\right)^{-1} \hv_{j,j,k}|^2}{\hv_{j,j,k}^{\herm} \left(\Av_j +  \lambda_j \Iv\right)^{-2} \hv_{j,j,k}} \nonumber \\
\label{eq:desired_deno}
&=& \frac{|\hv_{j,j,k}^{\herm} \left(\Av_{jk} + \lambda_j \Iv\right)^{-1} \hv_{j,j,k}|^2}{\hv_{j,j,k}^{\herm} \left(\Av_{jk} +  \lambda_j \Iv\right)^{-2} \hv_{j,j,k}}
\eea
where $\Av_j = \sum_{(m,n)} \beta_{m,n} \hv_{j,m,n} \hv_{j,m,n}^{\herm}$
, $\Av_{jk} = \sum_{(m,n) \neq (j,k)} \beta_{m,n} \hv_{j,m,n} \hv_{j,m,n}^{\herm}$
and the last equality comes from the Sherman-Morrison matrix inversion lemma.

First,
applying Theorem 3.4 in \cite{Debbah:11} to the term $\hv_{j,j,k}^{\herm} \left(\Av_{jk} + \lambda_j \Iv\right)^{-1} \hv_{j,j,k}$
in the numerator of (\ref{eq:desired_deno}) yields
\bea
\hv_{j,j,k}^{\herm} \left(\Av_{jk} + \lambda_j \Iv\right)^{-1} \hv_{j,j,k} -
\frac{\epsilon_{j,j,k}}{N_t} \trace \left(\bold{R}\Sigmab_{jk}\right) \xrightarrow{a.s.} 0
\eea
where $\Sigmab_{jk} = \left(\sum_{(m,n) \neq (j,k)} \frac{\beta_{m,n}}{N_t} \hv_{j,m,n} \hv_{j,m,n}^{\herm} + \frac{\lambda_j}{N_t} \Iv\right)^{-1}$.
By employing Theorem 1 in \cite{Wagner:12},
it follows
\bea \label{eq:m_det}
\hv_{j,j,k}^{\herm} \left(\Av_{jk} + \lambda_j \Iv\right)^{-1} \hv_{j,j,k} - m_{j,k}^{\circ}
\xrightarrow{a.s.} 0
\eea
where $m_{j,k}^{\circ} = \epsilon_{j,j,k} \trace\left(\bold{R}\phi(\Lc_{jk}, \frac{\lambda_j}{N_t})\right)$
and $\Lc_{jk} = \Lc_j \backslash \{\epsilon_{j,j,k}\beta_{j,k}\}$ with $\Lc_j = \{\epsilon_{j,1,1} \beta_{1,1} , \epsilon_{j,1,2} \beta_{1,2}, \ldots$ $, \epsilon_{j,M,K} \beta_{M,K} \}$.
Here, $\phi(\Sc,\rho)$ is defined as
\bea
\label{def:Tv_k}
\phi(\Sc,\rho) = \left(\frac{1}{N_t}\sum_{s_i\in\mathcal{S}} \frac{s_i\bold{R}}{1+e_i}+\rho\bold{I}\right)^{-1}
\eea
where $\Sc$ equals a set with non-negative elements $s_i$ for $i = 1, \ldots, |\Sc|$, $\rho$ represents a positive scalar value
and $e_i$'s are unique positive solutions of the {fixed-point equations}
\bea
\label{def:e_i}
e_i = s_i\trace \left(\bold{R} \phi(\Sc,\rho)\right).
\eea

Next, for the denominator in (\ref{eq:desired_deno}),
Theorem 1 in \cite{Wagner:12} leads to
\bea
{\hv_{j,j,k}^{\herm} \left(A_{jk} + \lambda_j \Iv \right)^{-2} \hv_{j,j,k}} - \frac{\epsilon_{j,j,k}}{N_t^2} \trace \left(\bold{R}\Sigmab_{jk}^2\right)
\xrightarrow{a.s.} 0.
\eea
Then, adopting Theorem 2 in \cite{Wagner:12},
we can write
\bea \label{eq:power_norm_det}
{\hv_{j,j,k}^{\herm} \left(A_{jk} + \lambda_j \Iv \right)^{-2} \hv_{j,j,k}} - \frac{1}{N_t}\Psi_{j,k}^{\circ} \xrightarrow{a.s.} 0
\eea
where $\Psi_{j,k}^{\circ} = \epsilon_{j,j,k} \trace\left(\bold{R}\phi^{'}(\Lc_{jk}, \frac{\lambda_j}{N_t})\right)$.
Here, $\phi^{'}(\Sc, \rho)$ is denoted as
\bea \label{def:Tv_derivate}
\phi^{'}(\Sc, \rho)
=  \phi(\Sc,\rho)\left( \bold{I} + \frac{1}{N_t} \sum_{s_i\in\mathcal{S}} \frac{s_i e_i^{'}\bold{R} }{(1 + {e_i})^2} \right) \phi(\Sc,\rho)
\eea
and $\ev^{'} = [e_1^{'}, \ldots, e_{|\Sc|}^{'} ]^{\transp}$ is expressed by
\bea
\ev^{'} = \left( \Iv_{|\Sc|} - \Jv \right)^{-1} \vv,\nonumber
\eea
where $\Jv$ and $\vv$ are computed as
\bea
[\Jv]_{ij} &=& \frac{ s_i s_j \trace \left(\bold{R} \phi(\Sc, \rho)\bold{R} \phi(\Sc, \rho)\right)}{N_t (1 + e_j)^2} \qquad \mbox{  for } i,j = 1, \ldots, |\Sc|   \nonumber \\
 \vv &=& \left[s_1\trace \left(\bold{R} \phi(\Sc,\rho)^2\right) , \ldots, s_{|\mathcal{S}|}\trace \left(\bold{R} \phi(\Sc, \rho)^2\right) \right]^{\transp}.\nonumber
\eea
Finally, combining (\ref{eq:power_norm_det}) and (\ref{eq:m_det}), we have
\bea
|\hv_{j,j,k}^{\herm} \bar{\vv}_{j,k}|^2 - { \frac{(m_{j,k}^{\circ})^2}{\frac{1}{N_t} \Psi_{j,k}^{\circ}}} \xrightarrow{ a.s.} 0.
\eea

\subsubsection{Deterministic equivalent for $|\hv_{j,m,n}^{\herm} \bar{\vv}_{j,k}|^2$}
The interference term $|\hv_{j,m,n}^{\herm} \bar{\vv}_{j,k}|^2$ can be written as
\bea
|\hv_{j,m,n}^{\herm} \bar{\vv}_{j,k}|^2
\!\!\!&=& \frac{|\hv_{j,m,n}^{\herm} \left(\Av_j + \lambda_j \Iv\right)^{-1} \hv_{j,j,k}|^2}{\|\left(\Av_j + \lambda_j \Iv\right)^{-1} \hv_{j,j,k}\|^2} \nonumber \\
\!\!\!&=& \frac{\hv_{j,m,n}^{\herm} \left(\Av_{jk} + \lambda_j \Iv\right)^{-1} \hv_{j,j,k} \hv_{j,j,k}^{\herm}\left(\Av_{jk} + \lambda_j \Iv\right)^{-1}\hv_{j,m,n}}
{\hv_{j,j,k}^{\herm} \left(\Av_{jk} + \lambda_j \Iv\right)^{-2} \hv_{j,j,k}} \nonumber \\
\label{eq:int_sig}
\!\!\!&=& \frac{\hv_{j,m,n}^{\herm} \left(\Av_{jkmn} + \lambda_j \Iv\right)^{-1} \hv_{j,j,k} \hv_{j,j,k}^{\herm}\left(\Av_{jkmn} + \lambda_j \Iv\right)^{-1}\hv_{j,m,n}}
{\left(1+ \beta_{m,n} \hv_{j,m,n}^{\herm}\left(\Av_{jkmn} + \lambda_j \Iv\right)^{-1}\hv_{j,m,n}\right)^2 \left(\hv_{j,j,k}^{\herm} \left(\Av_{jk} + \lambda_j \Iv\right)^{-2} \hv_{j,j,k}\right)}
 \eea
where $\Av_{jkmn} = \sum_{(i,q) \neq (j,k),(m,n)} \beta_{i,q} \hv_{j,i,q} \hv_{j,i,q}^{\herm}$ and the second and third equality come from the Sherman-Morrison matrix inversion lemma with respect to $\hv_{j,j,k}$ and $\hv_{j,m,n}$, respectively.

First, by applying Theorem 3.4 in \cite{Debbah:11} to the numerator in (\ref{eq:int_sig}) twice with respect to $\hv_{j,j,k}$ and $\hv_{j,m,n}$,
we can easily show that
\bea
\hv_{j,m,n}^{\herm} \left(\Av_{jkmn} + \lambda_j \Iv\right)^{-1} \hv_{j,j,k} \hv_{j,j,k}^{\herm}\left(\Av_{jkmn} + \lambda_j \Iv\right)^{-1}\hv_{j,m,n} - \frac{\epsilon_{j,j,k} \epsilon_{j,m,n}}{N_t^2} \trace \left(\bold{R}^2\Sigmab_{jkmn}^2 \right) \xrightarrow{a.s.} 0
\eea
where $\Sigmab_{jkmn}$ is defined as $\Sigmab_{jkmn} \!\!=\!\!  \left(\frac{\lambda_j}{N_t} \Iv_{N_t}  \!+\! \sum_{(i,q) \neq (j,k),(m,n)} \frac{\beta_{i,q}}{N_t} {\hv_{j,i,q}\hv_{j,i,q}^{\herm}}\right)^{-1}$.
From Theorem 1 in \cite{Wagner:12},
it follows
\bea
\frac{\epsilon_{j,j,k} \epsilon_{j,m,n}}{N_t^2} \trace\left(\bold{R}^2 \Sigmab_{jkmn}^2\right) - \frac{1}{N_t}\Psi_{j,k,m,n}^{\circ} \xrightarrow{a.s.} 0
\eea
where $\Psi_{j,k,m,n}^{\circ} = \epsilon_{j,j,k} \epsilon_{j,m,n} \trace\left(\bold{R}^2\phi^{'}(\Lc_{jkmn}, \frac{\lambda_j}{N_t})\right)$ with $\Lc_{jkmn} = \Lc_{jk}\backslash\{\epsilon_{j,m,n}\beta_{m,n}\}$.

Similarly, in the denominator of (\ref{eq:int_sig}),
one can show that
\bea
{\hv_{j,m,n}^{\herm}}  \left(\Av_{jkmn} + \lambda_j \Iv\right)^{-1} {\hv_{j,m,n}} -  m_{j,k,m,n}^{\circ} \xrightarrow{a.s.} 0
\eea
where $m_{j,k,m,n}^{\circ} = \epsilon_{j,m,n} \trace\left(\bold{R}\phi( \Lc_{jkmn} , \frac{\lambda_j}{N_t})\right)$.
Also, from (\ref{eq:power_norm_det}), we know that the second term converges almost surely to $\frac{1}{N_t}\Psi_{j,k}^{\circ}$.
Then, combining all results generates
\bea
|\hv_{j,m,n}^{\herm} \bar{\vv}_{j,k}|^2 - { \frac{\Psi_{j,k,m,n}^{\circ}}{\left(1+ \beta_{m,n} m_{j,k,m,n}^{\circ}\right)^2 \Psi_{j,k}^{\circ}}} \xrightarrow{ a.s.} 0,
\eea
and this concludes the proof.

\bibliographystyle{IEEEtran}

\end{document}